\documentclass[11pt]{article}
\usepackage{amsfonts}
\usepackage{amscd}
\usepackage{amsmath}
\newlength{\dinwidth}
\newlength{\dinmargin}
\makeatletter
\@addtoreset{equation}{section}
\makeatother

\def\be{\begin{equation}}
\def\ee{\end{equation}}
\def\ben{\begin{displaymath}}
\def\een{\end{displaymath}}
\def\baa{\begin{eqnarray}}
\def\eaa{\end{eqnarray}}

\def\ba{\begin{array}}
\def\ea{\end{array}}

\def\a{\alpha}
\def\g{\gamma}

\def\b{\beta}

\def\l{\lambda}

\def\th{\vartheta}
\def\Th{\Theta}

\def\O{\Omega}
\def\eb{{\bf e}}
\def\UC{{\hat{T}}}
\def\lp{x}
\def\CP1{{\mathbb C}P^1}
\def\Pcal{{\cal P}}
\def\Sh{\widehat{S}}
\def\Thpq{\Th\left[^\pb_\qb\right]}

\def\B{{\bf B}}
\def\C{{\mathbb C}}

\def\la{\label}

\def\f{\frac}
\def\L{{\cal L}}
\def\p{\partial}
\def\pb{{\bf p}}
\def\qb{{\bf q}}

\def\zb{{\bf z}}
\def\bk{{\bf k}}
\def\tr{{\rm tr}}
\def\0{S}
\def\log{\ln}

\def\w{{\bf w}}
\def\M{{\cal M}}

\def\B{{\bf B}}
\def\la{\label}

\def\f{\frac}
\def\L{{\cal L}}
\def\p{\partial}
\def\res{{\rm res}}

\def\tr{{\rm tr}}
\def\0{S}
\def\1{T}

\def\log{\ln}

\def\bar{\overline}
\def\det{{\rm det}}
\def\dbar{\bar{\partial}}

\newtheorem{condition}{Condition}
\newtheorem{remark}{Remark}
\newtheorem{definition}{Definition}
\newtheorem{theorem}{Theorem}
\newtheorem{corollary}{Corollary}
\newtheorem{lemma}{Lemma}

\def\Box{\diamond}

\begin{document}

\title{Solution of matrix Riemann-Hilbert problems with
quasi-permutation monodromy matrices}
\author{
D. Korotkin
\footnote{e-mail: korotkin@mathstat.concordia.ca}}

\maketitle

\begin{center}
Department of Mathematics and Statistics, Concordia University\\
7141 Sherbrooke West, Montreal H4B 1R6, Quebec,  Canada
\end{center}

\vskip0.5cm
{\bf Abstract.}
In this paper we solve an arbitrary matrix  Riemann-Hilbert (inverse monodromy) problem with
quasi-permutation monodromy representations
 outside of a
divisor in the space of monodromy data. This divisor is characterized in terms
 of the theta-divisor on the Jacobi manifold of  an auxiliary compact
Riemann surface realized as an appropriate branched
 covering of $\CP1$ . The solution is given in 
terms of a generalization of  Szeg\"o kernel
on the  Riemann surface. In particular, our construction provides a new  class of solutions of the Schlesinger system.
The isomonodromy tau-function of these solutions is computed up to a
nowhere vanishing factor independent of the elements of monodromy matrices. 
Results of this work generalize the results of 
 papers \cite{KitKor98} and \cite{DIKZ98} where the $2\times 2$ case
 was solved.

\vskip1.0cm

{\it subjclass:} {Primary 35Q15; Secondary 30F60, 32G81.}

\section{Introduction}

Apart from pure mathematical significance (see review of  A.Bolibruch \cite{Bolibruch}),
matrix Riemann-Hilbert (inverse monodromy) problems and related theory of isomonodromic deformations
play an important role in  mathematical physics.
In particular, the RH problems 
are central in the theory of integrable systems
(see for example \cite{ZMNP80,Dub94,Hitc94}) and the theory of random matrices \cite{Deift}.
In applications  the main object of interest is the so-called tau-function, which was first introduced 
by M.Jimbo, T.Miwa and their collaborators \cite{JimMiw81}; it was later shown by B.Malgrange \cite{Malg83} that the
tau-function may be interpreted as  determinant of certain T\"oplitz operator. The set of
zeros of the tau-function in the space of singularities of the RH problem is called 
the Malgrange divisor $(\th)$; it plays a crucial role in discussion of solvability
of RH problem with given monodromy data.

For generic monodromy data neither the solution  of a matrix RH problem
nor the corresponding tau-function  can  be computed analytically in terms
of known special functions \cite{Umem90,Wata99}. However, there are exceptional cases, when the RH
problem can be solved explicitly; surprisingly enough, these cases often appear in applications.
For example,  the solution of $2\times 2$ RH problem with an arbitrary set of off-diagonal
monodromy matrices was successfully  applied to the problem of finding physically meaningful
solutions of stationary axially symmetric Einstein equations \cite{Koro88,NeuMei,Klein} and to complete classification of 
$SU(2)$-invariant self-dual Einstein manifolds \cite{Hitc94,BabKor98}. The solution of general 
$2\times 2$ RH problem of this kind was given only in 1998 in the papers  \cite{KitKor98,DIKZ98}
(however, some important ingredients of this solution were understood
already  three decades ago, see review \cite{Zver71}). 
In \cite{KitKor98} it was
also calculated the tau-function corresponding to this RH problem, which turned out to coincide with 
determinant of Cauchy-Riemann operator acting in  tensor product of the
spinor bundle and an appropriate flat line bundle on a  hyperelliptic curve 
(see \cite{Zamo86,BelKni87,Kniz87,AlMoVa86}). 
In \cite{Koro00} a family of  Riemann-Hilbert problems in arbitrary
matrix dimension
with  quasi-permutation monodromies was solved in terms of
Szeg\"o kernel on compact Riemann surfaces; however, this family did
not contain enough parameters to cover the whole set of
quasi-permutation monodromy groups; also the Miwa-Jimbo tau-function
was not computed for dimension higher than 2.

Results of present work generalize the results of 
 papers \cite{KitKor98,DIKZ98} and \cite{Koro00}; we present a
 complete solution of Riemann-Hilbert problems with an 
arbitrary  quasi-permutation monodromy representation in
any matrix dimension
 outside of the  divisor of zeros of corresponding tau-function in the space of monodromy data
(by monodromy data we mean the given monodromy representation and positions of singularities).
For that purpose we use an appropriate
 generalization of  Szeg\"o kernel on associated Riemann surface. This leads to a new
class of solutions of the Schlesinger system.
We compute the Jimbo-Miwa tau-function up to a factor which depends
only on positions of singularities of the RH problem and does not
depend on the matrix elements of monodromy matrices; in some 
 cases (for matrix dimension $2$ and for RH problems in arbitrary
matrix dimension corresponding to Riemann surfaces of genus
$0$ and $1$) this factor can also be found explicitly. 
From the point of view of string theory \cite{Kniz87} this factor can in some cases be interpreted as 
determinant of Cauchy-Riemann operator acting in trivial line bundle over $\L$;
from the point of view of the theory of Frobenius manifolds this factor is equal to
isomonodromic tau-function of Frobenius manifolds associated to Hurwitz spaces \cite{KokKor03}.
The  divisor of zeros
of the tau-function corresponding to our RH problem in the space of monodromy data can be 
characterized in terms
 of the theta-divisor on the Jacobi manifold of the Riemann  surface.

The main technical tools used in this paper are kernel functions on Riemann surfaces, Fay identities
and deformation theory of Riemann surfaces. The systematic description of these objects may be found
in Fay's books \cite{Fay73,Fay92}. 

We expect present results to find an application to  the problem of isolating the subclass of physically
reasonable solutions of stationary axially symmetric Einstein-Maxwell system \cite{Koro88} in the
spirit of works \cite{Koro88,NeuMei,Klein}, devoted to vacuum Einstein equations. For Einstein-Maxwell system the 
 matrix dimension of RH problem is equal to  three.
Other potential areas of application are the theory of Frobenius manifolds \cite{Dub94} and random
matrices \cite{Deift}. 

Let's describe the  organization of this paper.
In section 2 we remind the formulation of general Riemann-Hilbert (inverse monodromy problem),
the isomonodromy deformation equations (Schlesinger system), and definition  of Jimbo-Miwa tau-function.
We further 
discuss  quasi-permutation monodromy representations and their natural
relationship to  branched coverings of $\CP1$.

In section 3  we review the necessary facts from the deformation theory of Riemann surfaces and
adjust them to the situation when the Riemann surface is realized as a branched covering of the
complex plane. 

In section 4 we  solve an arbitrary RH problem with irreducible
quasi-permutation monodromy representation
outside of a divisor in the space of monodromy data.

In section 5 we describe corresponding solutions of Schlesinger system.

Section 6 is devoted to computation of  corresponding tau-function;  the
divisor of the zeros of tau-function is described
 in terms of theta-divisor on Jacobi manifold  of an auxiliary branch covering.

\section{Riemann-Hilbert  problem with quasi-permutation monodromies
and branched coverings of $\CP1$}

\subsection{Riemann-Hilbert problem, isomonodromy deformations and tau-function}

Consider a set of 
$M+1$ points $\l_0,\l_1,\dots,\l_M\in\C$, and a given $GL(N)$
monodromy representation $\M$ of
$\pi_1[\CP1\setminus\{\l_1,\dots,\l_M\}]$. Let us formulate the following Riemann-Hilbert problem:

{\it Find function $\Psi(\l)\in GL(N,\C)$, defined on universal cover of $\CP1\setminus\{\l_1,\dots,\l_M\}$,
which satisfies the following conditions: }
\begin{enumerate}\it
\item
 $\Psi(\l)$ is normalized at a point $\l_0$ on some sheet of the universal cover as follows:
\be
\Psi(\l_0)=I\;;
\la{norm}
\end{equation}
\item
$\Psi(\l)$  has given right holonomy $\M_\g$ along each  contour $\g\in\pi_1[\CP1\setminus\{\l_1,\dots,\l_M\}]$;
\item
$\Psi(\l)$  has regular singularities at the points $\l_n$ (i.e. function $\Psi$ 
grows at a neighbourhood of $\l_m$ not faster than some power of $\l-\l_m$).
\end{enumerate}

Consider the following set of standard generators $l_1,\dots,l_M$ of
$\pi_1[\CP1\setminus\{\l_1,\dots,\l_M\}]$. Choose  $\l_0$ to be the starting point and assume that the contour $l_n$
starts and ends at $\l_0$ such that the interior of $l_n$ contains only one marked point $\l_n$
(our convention is that the point $\l=\infty$ belongs to the exterior of any closed contour on $\CP1$). 
Moreover, we  assume that  these generators are ordered according to the following relation:
\be
l_M l_{M-1}\dots l_1 = {\bf 1}\;.
\la{rel}\ee 
The matrices $\M_{\g_n}:= \M_n$ are called monodromy matrices. 
As a corollary of 
(\ref{rel})  we have:
\be
\M_{M}\M_{M-1}\dots \M_{1} = I\;.
\la{Moninf}
\ee

We shall consider only the monodromy groups for which the singularity
of solution $\Psi$ of the RH problem at the points $\l_n$ has the following form:
\be
\Psi(\l)= \{G_n+ O(\l-\l_n)\}(\l-\l_n)^{T_n} C_n\;,
\la{regsing}\ee
where $G_n, C_n\in GL(N)$; $T_n={\rm diag}(t^{(1)}_n,\dots t^{(N)}_n)$. 

The monodromy matrices $\M_{n}$ are in this case related to coefficients of asymptotics (\ref{regsing}) as follows:
\be
\M_{n}= C_n^{-1} e^{2\pi i T_n} C_n\;.
\la{MnTn}
\ee 
i.e. all these matrices are diagonalizable (of course, not
simultaneously in non-trivial cases).
The set   $\{\l_n, \;\M_{n}\;,\; T_n\;,\; n=1,\dots,M\}$ is called the set of monodromy data. 

Solution  $\Psi(\l)$ of such RH problem  satisfies the following  matrix differential equation with meromorphic coefficients with simple poles:
\be
\f{d\Psi}{d\l}=\sum_{n=1}^M \f{A_n}{\l-\l_n}\Psi\;,
\la{eql}\ee
where
\be
A_n= G_n T_n G_n^{-1}\;.
\la{An}\ee
Suppose now that matrices $C_n$ and $T_n$ (and, therefore, the monodromy matrices) don't depend on positions of singularities  $\{\l_n\}$. Then function $\Psi$, in addition to (\ref{eql}),
 satisfies  the equations with respect to positions of singularities $\l_n$:
\be
\f{d\Psi}{d\l_n}=\left(\f{A_n}{\l_0-\l_n}-\f{A_n}{\l-\l_n}\right)\Psi\;.
\la{eqln}\ee
Compatibility conditions of equations (\ref{eql}) and (\ref{eqln}) imply 
  Schlesinger equations for residues $A_n$:
\ben
\f{\p A_n}{\p \l_m}= \f{[A_n,\,A_m]}{\l_n-\l_m} - \f{[A_n,\,A_m]}{\l_0-\l_m}\;,\hskip0.6cm
m\neq n\; ;
\een
\be 
\f{\p A_m}{\p \l_m}= -\sum_{n\neq m}\left(\f{[A_n,\,A_m]}{\l_n-\l_m} - 
\f{[A_n,\,A_m]}{\l_n-\l_0}\right)\;.
\la{Schl}\ee
Once a solution of the Schlesinger system is given, one can define the  tau-function \cite{JimMiw81}
by the system of equations
\be
\f{\p}{\p\l_n}\log\tau = H_n:= \f{1}{2}{\rm res}|_{\l=\l_n}\tr\left(\Psi_\l\Psi^{-1}\right)^2\;;
\hskip0.8cm 
\f{\p\tau}{\p\bar{\l_n}} = 0\;.
\la{taudef}\ee

According to Malgrange \cite{Malg83}, the isomonodromic tau-function can be interpreted as determinant
of certain T\"oplitz operator. 
The important role in the theory of RH problems is played by the divisor of zeros of the tau-function in the universal covering of the space 
$\{\{\l_m\}\in \C^M\,\big|\;
\l_m\neq\l_n\;\;if \;\; m\neq n\}$. In analogy to the  theta-divisor $(\Th)$ on a Jacobi variety, Malgrange denoted this divisor by $(\th)$.
The divisor $(\th)$ has the  following meaning: if  $\{\l_n\}\in (\th)$,
the Riemann-Hilbert problem with the given set of monodromy matrices and eigenvalues $t_n^{(j)}$
does not have a solution; the solution $\{A_m\}$ of Schlesinger system is singular on $(\th)$.

\subsection{Quasi-permutation monodromy representations and branched coverings}

In this paper we shall consider  two special kinds of $N\times N$
monodromy representations.
\begin{definition}\la{permut}
Representation $\M$ is called  the permutations representation if matrix $\M_{\g}$ is a  permutation matrix
for each $\g\in\pi_1[\CP1\setminus\{\l_1,\dots,\l_M\}]$. 
\end{definition}
Remind that a matrix is called  the matrix of permutation if each raw and each column of this matrix
contain exactly one non-vanishing entry and this entry equals to 1. 
Permutation matrices of size $N\times N$ are in natural  one-to-one correspondence with elements of permutation group $S_N$.
The definition (\ref{permut}) is self-consistent since the
product of any two permutation matrices  is again a permutation matrix.

\begin{theorem}\la{repcurve}
There exists a one-to-one correspondence between  $N\times N$ permutation representations of
$\pi_1[\CP1\setminus\{\l_1,\dots,\l_M\}]$ and  compact Riemann
surfaces (not necessarily connected) realized as $N$-fold ramified coverings of $\CP1$ with
projections of branch points  on $\CP1$ equal to $\l_1,\dots,\l_M$.  
\end{theorem}
{\it Proof.}
Given a ramified covering 
$\L$ with projections $\l_1,\dots,\l_M$ of branch points on $\CP1$, we
construct the corresponding permutation representation as follows. 
Denote the projection 
of $\L$ on $\CP1$ by $\Pi$.
Generators $\M_n$ of permutation monodromy group are given by the
following construction. Consider the 
lift  $\Pi^{-1}(l_n)$ of the generator
$l_n$ on $\L$.
This  is a union of $N$ (not necessary closed) non-intersecting contours on $\L$ which start and end at
some of the
points $\l_0^{(j)}$ (by $\l^{(j)}$ we  denote the point of $j$th sheet of  $\L$ 
which  has projection $\l$ on $\CP1$). Denote by $l_n^{(j)}$ the component of 
$\Pi^{-1}(l_n)$ which starts at the point $\l_0^{(j)}$; the endpoint of this
contour is $\l_0^{(j_n[j])}$ for some index $j_n[j]$.
If $\l_n^{(j)}$ is not a branch point, then $j_n[j]=j$, and contour $l_n^{(j)}$ is closed; if $\l_n^{(j)}$ is a branch point, then 
$j_n[j]\neq j$ and contour $l_n^{(j)}$ is non-closed.
Then the permutation  matrix $\M_n$ has 
the following form:
\be
\left(\M_{n}\right)_{jl} =\delta_{j_n[j], l}
\la{Mperm}
\ee
and naturally  corresponds to some element $s_n$ of the permutation group $S_N$.
On the other hand, starting from some permutation monodromy representation we can  
glue $N$ copies of $\CP1$ at the  branch points $\{\l_n\}$ in such a way that
 the obtained compact Riemann surface corresponds to  the permutation
monodromies (\ref{Mperm}) (see \cite{GriHar}, p.257).

$\Box$


\begin{definition}
Representation $\M$  is called the  quasi-permutations 
representation if $\M_\g$ is a quasi-permutation matrix for any $\g\in\pi_1[\CP1\setminus\{\l_1,\dots,\l_M\}]$.
\end{definition}
Again, this definition is natural since all quasi-permutation matrices form a subgroup in $GL(N)$. 
Remind  that a  matrix is called the   quasi-permutation matrix if each raw and each column of this matrix
contain only one non-vanishing entry.

We shall call two quasi-permutation representations $\M$ and ${\M}'$ equivalent if 
there exists some {\it diagonal} matrix $D$ with ${\rm det} D=1$ such that
\be
{\M}'_{\g}=D \M_{\g} D^{-1}
\la{equiv}\ee
for all $\g\in \pi_1[\CP1\setminus\{\l_1,\dots,\l_M\}]$.

To every quasi-permutation representation $\M$ we can naturally assign a
permutation representation ${\M}^0$ substituting 
$1$ instead of all 
non-vanishing entries  of all monodromy matrices; then from ${\M}^0$ we reconstruct 
the branched covering $\L$.

We shall consider quasi-permutation monodromy representations $\M$ which satisfy the following
additional conditions:
\begin{condition}\la{product}
Representation $\M$ can not be decomposed into direct sum of  two other 
representations, both of whose are quasi-permutation representations with respect to the
same basis in $\C^N$.
\end{condition}

\begin{condition}\la{nondiag}
Monodromy matrices of representation $\M$ can not be simultaneously diagonalized.
\end{condition}

Condition \ref{product} obviously implies that the permutation representation ${\M}^0$ 
also can not be decomposed into a direct sum of two representation both of which are
permutation representations in the same basis; in turn, this implies connectedness of
corresponding branched covering $\L$. The condition \ref{product} is weaker than the standard condition of irreducibility of $\M$: there exist reducible quasi-permutation representations which are however  irreducible into a product of two quasi-permutation representations (for example, any permutation representation is reducible 
in usual sense since there
exists an invariant subspace $x_1+\dots +x_N=0$).

Condition \ref{nondiag} is imposed for convenience: it guarantees that the matrix
 Riemann-Hilbert problem is not trivially reducible to $N$ independent scalar Riemann-Hilbert problems.

\begin{definition}\la{QL}
Denote by ${\cal Q}(\L)$ the space of orbits  of
the group (\ref{equiv}) acting on the space of irreducible
quasi-permutation monodromy representations corresponding to a given
connected branched covering $\L$.
\end{definition}

\begin{lemma}\la{counting}
The manifold ${\cal Q}(\L)$ has dimension $MN-2N+1$; its
 universal covering $\widehat{{\cal Q}(\L)}$ 
is isomorphic to $\C^{MN-2N+1}$.
\end{lemma}
{\it Proof.}
Let us first prove that  ${\cal Q}(\L)$ is a
$MN-2N+1$-dimensional space. The space of $M-1$ quasi-permutation
matrices has dimension $(M-1)N$ (matrix $\M_M$ can be expressed in
terms of $\M_1,\dots,\M_{M-1}$ according to (\ref{Moninf}).
Let us prove that the orbits of the action  (\ref{equiv}) by diagonal
matrices $D$ are $N-1$-dimensional. 
Infinitesimally, matrix $D$ can be written as $D=I+\epsilon D_0$,
where $D_0$ is a traceless diagonal matrix; the action (\ref{equiv})
then takes the form $\M_\g\to \M_\g +[D_0,\M_\g]\epsilon$. If the
orbits have dimension less than $N-1$, there must exist a
non-vanishing diagonal traceless matrix $D_0$ commuting with all
$\M_\g$, which contradicts the condition \ref{nondiag}.

The  space  ${\cal Q}(\L)$ can be  covered by
$\C^{MN-2N+1}$ as follows
(the space ${\cal Q}(\L)$ is  non-simply-connected since
each monodromy matrix must
contain exactly $N$ non-vanishing entries).
Starting from an arbitrary $\M\in {\cal
Q}(\L)$, we define a point in
$\C^{(M-1)N}$,  whose coordinates  are equal to the
logarithms of non-vanishing entries of monodromy matrices
$\M_1,\dots,\M_{M-1}$ (i.e. the covering is defined by exponentiation
applied to each non-vanishing component).
The transformations (\ref{equiv}) act in this space as translations 
in $N-1$ independent directions; corresponding space of orbits is a
$MN-2N+1$-dimensional affine  space  which universally covers ${\cal Q}(\L)$.

$\Box$

Denote the branch points of $\L$ by $P_1,\dots, P_L$, where $L\geq M$;
the equality, $L=M$, takes place only 
if all branch points $P_k$ have different projections on $\l$-plane.
Denote the ramification indexes of the branch points  (i.e.  numbers of sheets glued at each point
$P_k$) by $\bk_1,\dots,\bk_L$ respectively.

\begin{remark}\rm
If some quasi-permutation monodromy matrix $\M_n$ is
diagonal, 
then corresponding matrix ${\M}^0_n$ is equal to 
$I$, and $\l_n$ is  not a projection of 
any branch point on $\CP1$.  However, in the sequel we shall treat such points  in the same
fashion as all other $\l_m$'s by assigning to all non-branch points
the ramification index 1. 
 All  formulas below
 are written in such form that this does not lead to any inconveniences or inconsistencies.
\end{remark}

\begin{lemma}\la{qpdia}
Every quasi-permutation matrix is diagonalizable.
\end{lemma}
The {\it proof} is simple: we can decompose $\C^N$ into a direct sum of orthogonal 
invariant subspaces such that in each subspace our quasi-permutation matrix acts as 
a power of elementary cyclic permutation multiplied by a diagonal matrix; it is
easy to verify that each such matrix has different eigenvalues and, therefore, is diagonalizable. Therefore, the original quasi-permutation  matrix acting in the whole
$\C^N$ is also diagonalizable.

$\Box$


According to the  Riemann-Hurwitz formula, the genus of  the connected Riemann surface $\L$ is equal to
\be
g=\sum_{l=1}^L\frac{\bk_l-1}{2} - N + 1\;.
\la{genus}\ee

Denoting the set of branch points by $\Pcal:=\{P_1,\dots,
P_L\}$, we get the  natural partition $\Pcal=\Pcal_1\cup\dots\cup\Pcal_M$,
where $\Pcal_m$ consists of  $s_m$  branch points $P_{s_1},\dots,P_{s_m}$ which project down
to $\l_m$ i.e. $\Pi(\Pcal_m)=\l_m$. Corresponding 
ramification indexes  $\{\bk_{s_1},\dots,\bk_{s_m}\}$ assigned to each $\l_m$
form the {\it passport} of the branch covering $\L$. The branched
coverings with fixed passport form a stratum of the Hurwitz space
$H_{g,N}$ of meromorphic functions of degree $N$ on Riemann surfaces
of genus $g$. The points $P_l$ are the critical points of these maps,
and $\l_m$ are corresponding critical values. The critical values
$\l_m$ can be used as local coordinates on the stratum of Hurwitz
space with given passport.

The stratum of highest dimension (i.e.  the bulk of the Hurwitz space) corresponds to branch
coverings with simple branch points (i.e. $\bk_m=2$ for all $m$).

\section{Riemann surfaces. Variational formulas}

\subsection{Riemann surfaces}

Here we collect some useful facts from the theory of Riemann surfaces and their 
deformations.
Consider a canonical basis of cycles  $(a_\a,b_\a),\;\a=1,\dots,g$ on $\L$. Introduce the dual basis of 
holomorphic 1-forms $w_\a$ on $\L$
normalized by $\oint_{a_\a}w_\b=\delta_{\a\b}$.  The matrix of $b$-periods $\B$ and the Abel map $U(P)\,,\;P\in\L$
are given by
\be
\B_{\a\b}=\oint_{b_\a} w_{\b}\;,\hskip0.7cm 
U_\a (P)=\int_{P_0}^P w_\a\;,
\la{Bw}\ee
where $P_0$ is a basepoint.
Consider theta-function with characteristics $\Thpq(\zb|\B)$, where $\pb,\qb\in\C^g$ are vectors of 
characteristics; $\zb \in\C^g$ is the argument. The theta-function is holomorphic function of variable $\zb$ with the
following periodicity properties:
\ben
\Thpq(\zb+\eb_\a)=   \Thpq(\zb) e^{2\pi i p_\a} \;;
\een
\be
\Thpq(\zb+\B\eb_\a)=   \Thpq(\zb) e^{-2\pi i q_\a}e^{-2\pi i z_\a-\pi i \B_{\a\a}}\;,
\la{perth}\ee
where $\eb_\a\equiv (0,\dots,1,\dots,0)$ is the standard basis in $\C^g$. 
The theta-function satisfies the heat equation:
\be
\f{\p^2\Thpq(\zb)}{\p z_\a\p z_\b}=4\pi i\f{\p\Thpq(\zb)}{\p\B_{\a\b}}\;.
\la{heat}\ee
Let us consider some non-singular odd half-integer characteristic $[\pb^*,\qb^*]$. 
The prime-form $E(P,Q)$ is defined as follows:
\be
E(P,Q)=\f{\Th\left[^{\pb^*}_{\qb^*}\right](U(P)-U(Q))}{h(P) h(Q)}\;,
\la{prime}\ee
where the square of a section $h(P)$ of a spinor bundle over $\L$ is given by the following expression:
\be
h^2(P)=\sum_{\a=1}^g \p_{z_\a}\left\{\Th\left[^{\pb^*}_{\qb^*}\right](0)\right\} w_a(P)\;.
\la{hp}\ee
Then  $h(P)$ itself is  a  section of the spinor bundle corresponding to characteristic
$[^{\pb^*}_{\qb^*}]$.  The automorphy factors of the prime-form along all cycles $a_\a$ are trivial;
the automorphy factor along  cycle $b_\a$ equals to  $\exp\{-\pi i B_{\a\a}- 2\pi i (U_\a(P)-U_\a(Q))\}$. 
The prime-form has the following local behavior as $P\to Q$:
\be
E(P,Q)=\frac{\lp(P)-\lp(Q)}{\sqrt{d\lp(P)}\sqrt{ d\lp(Q)}}(1+ o(1))\;,
\la{asprime}\ee
where $\lp(P)$ is a local parameter.

The Bergmann kernel is defined by the formula
$\w(P,Q)=d_P d_Q\log E(P,Q)$.
It has a double pole with  the following local behavior on the diagonal $P\to Q$:
\be
\w(P,Q)= \left\{\f{1}{(\lp(P)-\lp(Q))^2} + H(\lp(P),\lp(Q))\right\} d\lp(P) d\lp(Q)\;.
\la{defH}
\ee
where $H(\lp(P),\lp(Q))$ is the non-singular part of the Bergmann kernel in each coordinate chart. 
The restriction of the  function $H$ on the diagonal gives the Bergmann projective connection $R(\lp)$:
\be
R(\lp)= 6 H(\lp(P),\lp(P))\;,
\la{defR}\ee
which non-trivially  depends on the chosen system of local coordinates on $\L$.

The  Szeg\"o kernel $S(P,Q)$ is the $(1/2,1/2)$-form on $\L\times\L$  defined by the formula
\be
S(P,Q) = \f{1}{\Th\left[^\pb_\qb\right](0)}\f{\Th\left[^\pb_\qb\right](U(P)-U(Q))}{E(P,Q)}\;,
\la{szego}\ee
where $\pb,\qb\in\C^g$ are two vectors such that $\Th\left[^\pb_\qb\right](0)\neq 0$.
The Szeg\"o kernel is the  kernel of the integral  operator
$\bar{\partial}^{-1}$, where  the operator $\dbar$  acts in the line
bundle $\Delta\otimes \chi_{\pb,\qb}$, which is the product of the 
spin bundle (with trivial automorphy factors along the basic cycles) $\Delta$ over $\L$ 
(divisor of $\Delta$ is equivalent to vector of Riemann constants which we denote by the same letter) and the 
flat line bundle $\chi_{\pb,\qb}$  defined by
 the automorphy factors $e^{2\pi i p_\a}$ and $e^{-2\pi i q_\a}$ along basic cycles.
The Szeg\"o kernel itself  has the automorphy factors  $e^{2\pi i p_\a}$
and $e^{-2\pi i q_\a}$ along the cycles $a_\a$ and $b_\a$, respectively,
in its first argument; the automorphy factors of the Szeg\"o kernel with respect to its second
argument are the  inverse (i.e. $S(P,Q)$ is a section of the line bundle $\Delta\otimes \chi_{\pb,\qb}$ 
with respect to $P$ and a section of $\Delta\otimes \chi^{-1}_{\pb,\qb}$ with respect to $Q$). 
On the diagonal, as $Q\to P$, it behaves as follows:
\be
S(P,Q)=\left(\f{1}{x_P-x_Q}+ a_0(P) + O(x_P-x_Q)\right)\sqrt{d x_P}\sqrt{d x_Q}\;,
\la{Sdia}
\ee
where coefficient $a_0$ is given by (\cite{Fay92}, p.29)
\be
a_0(P)=\f{1}{d x_P}\sum_{\a=1}^g \p_{\a}\{\log\Thpq(0)\}{w_\a(P)}\;.
\la{a0P}
\ee 

The Szeg\"o kernel is related to the Bergmann kernel as follows (\cite{Fay73}, p.26):
\be
\hskip0.5cm - S(P,Q) S(Q,P) = \w(P,Q)+\sum_{\a,\b=1}^g\p^2_{z_\a z_\b}\{\log\Thpq(0)\}w_\a (P) w_\b(Q)\;.
\la{SzBerg}\ee

For any two sets   $P_1,\dots,P_N$ and $Q_1,\dots,Q_N$ of points on  $\L$
the following Fay identity takes place (see \cite{Fay73}, p.33):
\be\la{ident}
\det\{S(P_j,Q_k)\}
= \frac{\Th\left[^\pb_\qb\right]\left(\sum_{j=1}^N (U(P_j)-U(Q_j))\right)}
{\Th\left[^\pb_\qb\right](0)}\frac{\prod_{j<k} E(P_j,P_k) E(Q_k,Q_j)}{\prod_{j,k} E(P_j,Q_k)}\;.
\ee
In particular, for $N=2$ this is   Fay's trisecant identity.

\subsection{Rauch variational formulas}

The infinitesimal variation  of the basic  holomorphic 1-forms and
matrix of b-periods  with respect to a Beltrami differential $\mu$ is given by the 
Rauch  formulas (\cite{Fay92}, p.57):
\baa
\hskip0.3cm \delta_\mu w_\a(P)= \f{1}{2\pi i}\iint_{\L} \mu(Q) w_\a(Q) \w(P,Q)\;,\hskip0.8cm
 \bar{\delta}_\mu w_\a(P)= 0\;;
\la{Rauch1}\eaa
\be
\delta_\mu\B_{\a\b}=  \iint_{\L} \mu w_\a w_\b\;,\hskip0.8cm
\bar{\delta}_\mu\B_{\a\b}= 0\;.
 \la{Rauch2}\ee

Let us apply these formulas to a Riemann surface $\L$ realized as a
branched  covering of $\CP1$. 

\begin{theorem}
Basic holomorphic differentials and matrix of $b$-periods of an  $N$-fold covering $\L$ of
$\CP1$   satisfy the following equations:
 \be
\p_{\l_m} \{w_\a(P)\} = \res\Big|_{\l=\l_m} \left\{\f{1}{(d\l)^2}
\sum_{j} w_\a(\l^{(j)}) \w(\l^{(j)}, P)\right\}\;, 
\la{varw}
\ee
\be
\p_{\l_m}\{\B_{\a\b}\}= -\res\Big|_{\l=\l_m}\left\{\f{4\pi i}{(d\l)^2}\sum_{j< k}{w_\a(\l^{(j)})w_\b(\l^{(k)})}\right\}\;, 
\la{varB1}
\ee
\ben
\p_{\bar{\l}_m} \{w_\a(P)\}=
\p_{\bar{\l}_m}\B =0\;.
\een
where $\l^{(j)}$ denotes the point of $\L$ which has projection $\l$ on
$\l$-plane and belongs to the $j$th sheet of $\L$ (under certain dissection of
$\L$ into $N$ sheets). 
\end{theorem}
{\it Proof.}
We first notice that the residue in (\ref{varw}), (\ref{varB1}) and below is understood as
 the residue of 
function of variable $\l$, not the 1-form.
We start from proving   the theorem under assumption that all branch points of $\L$
have different projections on $\l$-plane i.e. there is a bijection
between the set of branch points $P_m$ and their projections $\l_m$;
then each $\Pcal_m$ contains only one point $P_m$. 

Choose in the Rauch formulas (\ref{Rauch1}), (\ref{Rauch2}) the
Beltrami differential as follows:
\begin{equation}
\mu=-\frac{1}{2\delta^{\bk_m}}\left(\frac{|\lp_m|}{\lp_m}\right)^{\bk_m-2}
{\bf  1}_{\{|\lp_m|\leq\delta\}}\frac{d\,\overline{\lp_m}}{d\,\lp_m}
\la{schif}
\end{equation}
with sufficiently small $\delta>0$, where  
$\lp_m\equiv(\l-\l_m)^{1/\bk_m}$ is a local parameter around
$P_m$; function ${\bf  1}_{\{|\lp_m|\leq\delta\}}$ is  equal to $1$
inside the   disc of radius $\delta$ centered at $\l_m$ and zero outside.  
If $\bk_m=2$, this is nothing but the
Schiffer variation; this variation acts on the moduli of the
Riemann surface in the same way as the delta-function with support at $P_m$.

Then formula (\ref{Rauch1}) gives rise  to (\ref{varw}).
Computing  the $b$-period of formula (\ref{varw}), we get
\be
\f{1}{2\pi i}\p_{\l_m}\{\B_{\a\b}\}=  \res\Big|_{\l=\l_m} \left\{\f{1}{(d\l)^2}\sum_{j=1}^N w_\a(\l^{(j)})w_\b(\l^{(j)})\right\}\;.
\la{varB}
\ee
In turn, this formula implies (\ref{varB1}) if we take into account the following lemma:

\begin{lemma} \la{holsum}
An arbitrary holomorphic differential $w(P)$ on a compact Riemann surface $\L$, realized as $N$-fold covering of $\CP1$, 
satisfies the following relation:
\be 
\sum_{j=1}^N w (\l^{(j)}) =0\;.
\la{sumw}\ee
\end{lemma}

{\it Proof.}
It is sufficient to check that $\sum_{j=1}^N w_\a (\l^{(j)})$ is a holomorphic differential on $\CP1$.
The suspicious points are the  branch points $P_m$.  Consider Taylor series of $w(P)$ in the neighborhood of 
the branch point:
$ w(\lp_m)=\sum_{n=1}^{\infty} A_n \lp_m^n d\lp_m $. We have to check the
regularity of the  expression 
\ben
\sum_{n=1}^{\infty}\left\{ \sum_{j=0}^{\bk_m-1}\gamma_m^{j(n+1)} \right\} A_n  \lp_m^n \frac{d\lp_m}{d\l}\,,
\een
where $\gamma=e^{2\pi i/\bk_m}$, in a  neighborhood of the point $P_m$. 
Taking into account that $d\l=\bk_m \lp_m^{\bk_m-1} d\lp_m$, this regularity follows from the fact that
$\sum_{j=0}^{\bk_m-1} \gamma^{j(n+1)}=0$ for $n=0,\dots,\bk_m-2$. 

$\Box$

Thus we proved the  formulas (\ref{varw}), (\ref{varB1}) for the case when
 each $\Pcal_m$ consists of only one branch  point.  Any family of
general coverings can be obtained in a smooth  limit from these
coverings if one assumes that some $\l_m$ coincide; the formulas
(\ref{varw}), (\ref{varB1}) are already written in the form which is
stable with respect to  such limiting procedure. 

$\Box$

We notice that the non-trivial contributions in variational  formulas (\ref{varw}),
(\ref{varB1}) arise only from the branch points contained in  $\Pi^{-1}(\l_m)$.

To write down variational formula for the Szeg\"o kernel it is
convenient to introduce function
\be
s(P,Q)=\f{S(P,Q)}{\sqrt{dx_P}\sqrt{dx_Q}}\;,
\la{sPQ}
\ee
which, obviously, depends on a choice of local parameters near points
$P$ and $Q$. We shall write down the variational formula for the
Szeg\"o kernel only in the partial case of simple branch points. For
the case of arbitrary multiplicities the variational formulas can be
deduced from the formulas of (\cite{Fay92}, p.56) similarly to
(\ref{varw}), (\ref{varB1}).

\begin{theorem}
Suppose that $\Thpq(0)\neq 0$.
Assume that all branch points of the covering $\L$ are simple and have
different projections on $\l$-plane, i.e. $\Pi(P_m)=\l_m$.
Suppose that the local parameters $dx_P$ and $dx_Q$ from (\ref{sPQ})
don't depend on (some) $\l_m$. Then
\be
\p_{\l_m}\{ s(P,Q)\}= \f{1}{4}\left\{ D_m [s(P,P_m)] s(P_m,Q)-
s(P,P_m) D_m [s(P_m,Q)]\right\}
\la{vars}
\ee
where
\be 
D_m[s(P,P_m)]:= \f{d s(P,Q)}{d x_m(Q)}\Big|_{Q=P_m}
\la{defDm}
\ee
\end{theorem} 

{\it Proof.} The formula  (\ref{vars}) can be deduced from general
variational  formula in
(\cite{Fay92}, p.56) by substitution of Schiffer variation. The
simple  independent  proof looks  as follows. The Szeg\"o kernel $S(P,Q)$ 
 behaves as follows as $P\to P_m$, when $x_P=\sqrt{\l-\l_m}$:
\ben
S(P,Q)= s(P,Q) \sqrt{d x_Q}\sqrt{d\sqrt{\l-\l_m}}\;,
\een
where 
$$
s(P,Q)= s(P_m,Q)+ D_m[s(P_m,Q)] \sqrt{\l-\l_m}+ O(\l-\l_m)\;.
$$
Differentiating 
$$\sqrt{d\sqrt{\l-\l_m}}=\f{\sqrt{d\l}}{\sqrt{2}(\l-\l_m)^{1/4}}$$
with respect to $\l_m$, we see that, as $P\to P_m$:
\ben
\p_{\l_m}\{S(P,Q)\} =\f{1}{4}\left\{\f{s(P_m,Q)}{\l-\l_m}-
\f{D_m[s(P_m,Q)]}{\sqrt{\l-\l_m}}\right\}\sqrt{d x_Q}\sqrt{d \sqrt{\l-\l_m}}\;.
\een

Analogous analysis of $\p_{\l_m}\{S(P,Q)\}$ as $Q\to P_m$ allows to
conclude that $\p_{\l_m}\{S(P,Q)\}$ has the same set of singularities
and singular parts as the expression
\be
\f{1}{4}\left\{s(P_m,Q)D_m[s(P, P_m)]- D_m[s(P_m,Q)]s(P,P_m)\right\}\sqrt{dx_P}\sqrt{dx_Q}
\la{derSPQ}
\ee
(the differentiation with  respect to $\l_m$ kills the pole of
$S(P,Q)$ as $P\to Q$). Moreover, $S(P,Q)$ and expression
(\ref{derSPQ}) are
meromorphic sections of the same
line 
bundle $\Delta\otimes \chi_{\pb,\qb}$ over $\L$ with respect to $P$ and
$\Delta\otimes \chi_{\pb,\qb}^{-1}$ with respect to $Q$. 
Since, as long as the Szeg\"o kernel is well-defined (i.e.  $\Thpq(0)\neq 0$), both of these bundles don't have
holomorphic sections (see for example \cite{Fay92}, p.29) we come to (\ref{vars}).

$\Box$

\section{Solution of Riemann-Hilbert problems with\\ quasi-permutation monodromies}

Here we solve Riemann-Hilbert problems with an  arbitrary 
 quasi-permutation monodromy representation $\M$ satisfying the 
non-triviality conditions \ref{product} and \ref{nondiag} and an arbitrary set of regular singularities, except a
divisor in the space of the monodromy data $\{\l_m,\;{\cal M}_m\}$.  Consider some
 quasi-permutation monodromy representation $\M$  
of $\pi_1[\CP1\setminus\{\l_1,\dots,\l_M\}]$ satisfying conditions \ref{product} and \ref{nondiag} and construct corresponding
 permutation representation ${\M}^0$; then construct the corresponding
 connected  branched
 covering $\L$ of $\CP1$. As before, denote the branch points of $\L$
 by $P_1,\dots, P_L$, where $L\geq M$, and their ramification indexes 
by $\bk_1,\dots,\bk_L$ respectively. The genus $g$ of $\L$ is given
 by the Riemann-Hurwitz formula (\ref{genus}). Introduce on $\L$ a canonical basis of cycles
$(a_\a, b_\a)$ such that the projections of the basic cycles on $\l$-plane don't pass through points
$\l_0,\l_1,\dots,\l_M$.

In the sequel it will be convenient to extend the notion of
ramification index $\bk_m^{(j)}$ to all of the
points $\l_m^{(j)}$  assuming that the ramification index  is equal to
$\bk_l$ if $\l_m^{(j)}$ coincides with the branch point $P_l$, and the
ramification index is equal to $1$ 
 if $\l_m^{(j)}$ is not a branch point.
Introduce the following set of parameters:
\begin{itemize}
\item
Two vectors $\pb,\qb\in \C^g$.
\item
Constants 
$r_m^{(j)}\in \C$ assigned to each point $\l_m^{(j)}$; we assume that the 
constants $r_m^{(j)}=r_m^{(j')}$ coincide if $\l_m^{(j)}=\l_m^{(j')}$ i.e. if $\l_m^{(j)}$ is
a branch point. We require that
\be
\sum_{m=1}^M\sum_{j=1}^N r_m^{(j)} =0\;.
\la{sumr}\ee
Therefore, among  constants $r_m^{(j)}$ we have 
\ben
MN-1 -\sum_{l=1}^L (\bk_l-1) = MN-2g-2N+1
\een
independent parameters naturally assigned to non-coinciding points among
$\l_m^{(j)}$. 
\end{itemize}

Hence, altogether we introduced $MN-2N+1$ independent constants
$\pb,\qb$ and $r_m^{(j)}$; according to lemma \ref{counting},
this number exactly equals the number of non-trivial parameters
carried by the  non-vanishing entries of  
the quasi-permutation 
monodromy matrices of our RH problem. 

Let us introduce on $\L$ a contour $S$, which connects some initial
point $P_0$ with  
all  points $\l_m^{(j)}$, including all branch points (we hope that the use of the same notation 
for this contour and the Szeg\"o kernel does not lead to a confusion).
Introduce also another  contour $S_0$, which connects the point $P_0$
only with the branch points of odd multiplicity (i.e. the branch
points with even ramification indexes $\bk_m$); the number of such branch points must be
even itself to get the integer genus via Riemann-Hurwitz formula.
We assume that both contours $S$ and $S_0$ don't intersect the basic
cycles, i.e. they belong to the interior of fundamental polygon $\hat{\L}$ of $\L$.

Suppose that the point $\l_0$ does not belong to the set of
projections of basic cycles $(a_\a,b_\a)$ and contours $S$ and $S_0$ on
$\CP1$. Let us define the 
intersection indexes of  the contours $l_m^{(j)}$ with all  basic cycles and the contour $S$:
\be
I_{m\a}^{(j)}=l_m^{(j)}\circ a_\a\;,\hskip0.6cm
J_{m\a}^{(j)}=l_m^{(j)}\circ b_\a \;,\hskip0.6cm
K_m^{(j)}= l_m^{(j)}\circ S \;,\hskip0.6cm
L_m^{(j)}= l_m^{(j)}\circ S_0 \;,
\la{inter}\ee
\ben
{\rm where}\;\;\;
m=1,\dots,M\;;\;\;\a=1,\dots,g\;;\;\;j=1,\dots,N\;.
\een
The contour $S$ can always be chosen in such a way that $K_m^{(j)}=1$ if $\l_m^{(j)}$ is not a branch point; if
$\l_m^{(j)}$ is a  branch point, then either $K_m^{(j)}=1$ or $K_m^{(j)}=0$.

Another auxiliary object we need to discuss is the lift of meromorphic spinor $\sqrt{d\l}$ from $\CP1$
to $\L$. On $\CP1$ this spinor has a single simple pole at $\l=\infty$. The 
differential $d\l$ on $\L$ has $N$ second order poles at all infinities $\infty^{(k)}$ 
and zeros of order $\bk_m-1$ at all branch points $P_m$. If all
ramification indexes $\bk_m$ are odd,
$\sqrt{d\l}$ must be a section of one of $4^g$ spinor bundles over
$\L$ (this case was discussed in \cite{Serr90}); if some of $\bk_m$ are even, $\sqrt{d\l}$ is a section of a
spinor bundle on $\L$ with additional
branch cut on the contour $S_0$, where  $\sqrt{d\l}$ changes its sign.

To find the half-integer characteristic $[\pb^0,\qb^0]$ which corresponds to
the spinor bundle defined  by  $\sqrt{d\l}$ we recall that the Abel map of the 
 divisor of spinor bundle with all
twists equal to $+1$ is equal to the vector of Riemann constants $\Delta$.
Therefore,  the difference between Abel map of
divisor of $\sqrt{d\l}$ and the vector of Riemann constants is equal to
$\B\pb^0+\qb^0$:
\be
\B\pb^0+\qb^0= \f{1}{2}\sum_{m=1}^L (\bk_m-1) U(P_m) - \sum_{j=1}^N U(\infty^{(j)}) -
\Delta 
\la{spintwist}
\ee
The automorphy factors of  $\sqrt{d\l}$ along cycles $a_\a$ and $b_\a$ are then
equal to $e^{2\pi i p^0_\a}$ and $e^{-2\pi i q^0_\a}$,
respectively.

Now we are in position to define the $N\times N$ matrix-valued function $\Psi(\l_0,\l)$
(we explicitly indicate dependence of $\Psi$ on normalization point $\l_0$ for future convenience) which will
later turn out to solve a Riemann-Hilbert problem. We  define the germ of function $\Psi(\l_0,\l)$
in a small neighborhood of the normalization point $\l_0$ by the following formula:
\be
\Psi_{kj}(\l_0,\l)=\psi(\l^{(j)},\l_0^{(k)})\;,
\la{psinew}\ee
where the scalar function $\psi(P,Q)$ ($P,Q\in \hat{\L}$) is defined by 
\be
\psi(P,Q)=
\Sh (P,Q)E_0(\l,\mu)\;,\hskip0.7cm \l=\Pi(P)\;,\hskip0.4cm \mu=\Pi(Q)\;,
\la{psismall}
\ee
and $\Sh(P,Q)$ is the modified Szeg\"o kernel, given by the following
formula inside of the fundamental polygon of Riemann surface $\L$:
\be
\Sh(P,Q):=\f{\Th\left[^\pb_\qb\right]\left(U(P)-U(Q)+\O\right)}
{\Th\left[^\pb_\qb\right](\O)E(P,Q)}\prod_{m=1}^M \prod_{l=1}^N \left[\f{E(P,\l_m^{(l)})}
{E(Q,\l_m^{(l)})}\right]^{r_m^{(l)}}\;.
\la{phinew}
\ee
By $E_0$ we denote the prime-form on $\CP1$
\ben
E_0(\l,\l_0)=
\frac{\l-\l_0}{\sqrt{d\l d\l_0}}\;,
\een
lifted to $\L$ as we discussed above;
\be
\O := \sum_{m=1}^M\sum_{j=1}^N r_m^{(j)} U(\l_m^{(j)})\;.
\la{Om}\ee
The vector $\O$ does not depend on the choice of initial point of the Abel map due to assumption
(\ref{sumr}). The formula (\ref{psinew}) makes sense 
 if $\Th\left[^\pb_\qb\right](\O)\neq 0$.

The following theorem gives a solution to  RH problems with 
quasi-permutation monodromies satisfying non-triviality conditions 
\ref{product} and \ref{nondiag} outside of divisor defined by equation $\Th\left[^\pb_\qb\right](\O)= 0$. This
is  the main result of this  section:

\begin{theorem}\la{Main}
Suppose that $\Th\left[^\pb_\qb\right](\O)\neq 0$.
Let us analytically continue function $\Psi(\l)$ (\ref{psinew}) from the neighborhood of the normalization point 
$\l_0$ to the universal covering $\UC$ of $\CP1\setminus\{\l_1,\dots,\l_M\}$.
Then the function $\Psi(\l)$ is non-singular and non-degenerate  on $\UC$. 
It  has regular singularities at the points $\l=\l_m$ of the form (\ref{regsing}), satisfies the 
normalization condition $\Psi(\l=\l_0)=I$ and 
solves the Riemann-Hilbert problem with  the following quasi-permutation monodromies:

\be\la{Mon}
\left(\M_n\right)_{jl}=\exp2\pi i\left\{ {\bk}_n^{(j)} r_n^{(j)}K_n^{(j)}
-\sum_{\a=1}^g \{J_{n \a}^{(j)}(p_\a+p_\a^0)+I_{n\a}^{(j)}
(q_\a+q_\a^0)\} +\f{1}{2}L_n^{(j)} \right\}\delta_{j_m[j],l}\;,
\ee
where all constants $\pb,\qb$ and $r_n^{(j)}$  
were introduced above; half-integer characteristic $[\pb^0,\qb^0]$ is
given by (\ref{spintwist}); the  intersection indexes are defined by
(\ref{inter}) ;
$j_m[j]$ stands for the number of the  sheet 
where the contour $l_m^{(j)}$ ends.
\end{theorem}

{\it Proof.}
Choose in the Fay identity  (\ref{ident}) $P_j= \l^{(j)}$ and $Q_k= \l_0^{(k)}$. Then, taking into 
account the holonomy properties of 
the prime-form and the asymptotics (\ref{asprime}), we conclude that
\ben
\det\Psi = \prod_{m=1}^M\prod_{j,k=1}^N \left[\f{E (\l^{(j)},\l_m^{(k)})}
{E(\l_0^{(j)},\l_m^{(k)})}\right]^{r_m^{(k)}}\;,
\een
 which, being considered as function of $\l$, does not vanish outside
of the points $\l_m^{(k)}$; thus $\Psi$
is non-degenerate and holomorphic if $\l$ does not coincide with any of $\l_m$.
The normalization condition $\Psi_{jk}\left(\l_0\right)=\delta_{jk}$ is an immediate corollary 
 of the asymptotic expansion of the  prime form (\ref{asprime}).

Expressions  (\ref{Mon}) for the monodromy matrices of function $\Psi$  follow from the simple 
consideration of the components of function $\Psi$. Suppose for a moment that the function $\Sh(P,\l_0^{(k)})E_0(\l,\l_0)$,
defined by (\ref{phinew}), would  be a single-valued function on $\L$ (as function of $P\in\L$). Then all monodromy
matrices would  be  matrices of permutation: the analytical continuation of  the matrix element 
 $\Sh(\l^{(j)},\l_0^{(k)})E_0(\l,\l_0)$ along contour $l_m^{(j)}$  would simply give the
matrix element $\Sh(\l^{(j_m[j])},\l_0^{(k)})E_0(\l,\l_0)$. However, since in fact the function 
$\Sh(P,\l_0^{(k)})E_0(\l,\l_0)$ gains some non-trivial multipliers from crossing the basic cycles $a_\a$, $b_\a$
and contour $S$, we get in (\ref{Mon}) an additional exponential factor.
Its explicit form is a corollary of  the definition of intersection indexes which enter this expression, and periodicity
properties of the  theta-function and the prime-form.

Function $\Psi$ is singular at the points $\l_m$ due to, first, the
product of the prime-forms in (\ref{phinew}), and, second, due to
different local parameters on $\l$-plane and on $\L$ which leads to
additional singularity in (\ref{psinew}). Obviously, the
singularity of $\Psi$ at $\l_m$ is regular.
$\Box$

To elucidate the role of constants $\pb,\,\qb$ and $r_m^{(j)}$ we shall compute the matrices $T_m$ from
(\ref{regsing}) which are the logarithms of the diagonal form of 
matrices ${\cal M}_m$ according to (\ref{MnTn}). For simplicity we consider the 
following ``model'' situation:

\begin{theorem}\la{Tm}
Suppose that ${\cal P}_m$ contains 
only one branch point $P_m$ and this branch point has degree $\bk_m$. Assume
that sheets number $j=1,\dots,\bk_m$ are glued at $P_m$ i.e. points $\l_m^{(j)}$ for $j=\bk_m+1,\dots, N$
are non-branch points.   Then
the elements of diagonal matrix $T_m$ are given by:
\be
t_m^{(j)} = r_m^{(j)}-\f{1}{2}+\f{1}{\bk_m}(j-\f{1}{2})\;,\hskip0.7cm j=1,\dots,\bk_m\;,
\la{tmj}
\ee
\be
t_m^{(j)} = r_m^{(j)}\;,\hskip0.7cm j=\bk_m+1,\dots,N
\la{tmj1}
\ee
(we recall that $r_m^{(1)}=\dots= r_m^{(\bk_m)}$). 
\end{theorem}

{\it Proof.}
To verify (\ref{tmj}) 
we first put all $r_m^{(j)}=0$. Then  the
singular part $(\l-\l_m)^{T_m}$  from (\ref{regsing}) of matrix $\Psi$ (\ref{psinew}) at $\l_m$
has the form
\be
\sqrt{\frac{d x_m}{d\l}} {\rm diag}(1,\,x_m,\dots, x_m^{\bk_m-1},1,\dots,1)\;,
\ee
where $x_m=(\l-\l_m)^{1/\bk_m}$, which leads to  (\ref{tmj}) with $r_m^{(j)}=0$ after computing
 ${d x_m}/{d\l}$. Coefficients $t_m^{(j)}$, $j=\bk_m+1,\dots,N$, which correspond to non-branch
 points $\l_m^{(j)}$ are in this case vanishing.

If we now introduce the non-trivial constants $r_m^{(j)}$, formulas (\ref{tmj1}) are obvious
from (\ref{psinew}), (\ref{psismall}), (\ref{phinew}).
To check (\ref{tmj}) we recall that, according to (\ref{psinew}), (\ref{phinew}), the  components  $\Psi_{kj}$ for $j=1,\dots,\bk_m$ get at $P_m$
an additional singularity of the form $x_m^{\bk_m r_m^{(j)}}$  i.e. $(\l-\l_m)^{r_m^{(j)}}$, which leads to (\ref{tmj}).

$\Box$

The form of matrix $T_m$ for an arbitrary branching structure over
$\l_m$ is a straightforward generalization of
(\ref{tmj}),(\ref{tmj1}).

\begin{theorem}
The theorem \ref{Main} provides the solution of the  
Riemann-Hilbert problem with an arbitrary set of singularities
$\{\l_m\}$ and an arbitrary  (up to equivalence (\ref{equiv})) 
 quasi-permutation monodromy representation, satisfying conditions
\ref{product} and \ref{nondiag},
outside of the divisor
in the $\{{\cal M}_m,\,\l_m\}$-space defined by the equation $\Th\left[^\pb_\qb\right](\O)=0$.
\end{theorem}

{\it Proof.}
Denote the vector space of dimension ${MN-2N+1}$ with coordinates
$p_{\a}, q_{\a}$ and $r_{m}^{(j)}$ by ${\cal H}$;
its $2g$-dimensional   subspace  defined by
equations  $r_{m}^{(j)}=0$ we  denote by  ${\cal H}_1$.  The
orthogonal subspace $\pb=\qb=0$ of dimension ${MN-2N+1-2g}$ is  denoted
by ${\cal H}_2$. 
The formulas (\ref{Mon}) for $\log\left\{({\cal M}_n)_{jl}\right\}$
define an affine map (denote it by ${\cal F}$) from ${\cal H}$ to the space  $\widehat{{\cal
Q}(\L)}$ (according to lemma 
\ref{counting},  its dimension is also equal  to ${MN-2N+1}$).

To show that an arbitrary  quasi-permutation monodromy representation
satisfying conditions \ref{product} and \ref{nondiag} is covered by 
the theorem \ref{Main} unless $\Th\left[^\pb_\qb\right](\O)=0$, it is sufficient to show 
that the affine map   ${\cal F}$ (\ref{Mon}) is non-degenerate.
Non-degeneracy of  ${\cal F}$ on ${\cal H}_2$
follows from the fact that,
according to proposition \ref{Tm},
constants ${r_{m}^{(j)}}$ determine the same number of different eigenvalues of monodromy matrices. 

On the other hand, vectors $\pb$ and $\qb$
don't enter the eigenvalues at all, i.e. these vectors influence only matrices $C_n$ 
from (\ref{MnTn}). Thus ${\cal F}({\cal H}_1)$ and ${\cal F}({\cal
H}_2)$ have only one common point - the image of the origin.
 Therefore, it remains to verify that the map 
${\cal F}$ is non-degenerate on ${\cal H}_1$.
The simplest way to verify this non-degeneracy is to observe that equivalent monodromy representations 
always  correspond to coinciding (up to a constant factor) isomonodromic tau-functions. As we
shall see below (\ref{tau00}), for $r_m^{(j)}=0$
the vectors $\pb$ and $\qb$ enter the tau-function only via
characteristics of the
 theta-function $\Thpq(0)$. This theta-function
obviously can not remain invariant on
any non-trivial linear subspace (independent  of $\{\l_m\}$) in the
$2g$-dimensional 
vector space spanned by the vectors $\pb$ and $\qb$.

$\Box$

\begin{remark}\rm
If we assume that all constants  $r_m^{(j)}$ vanish, the formula
(\ref{psinew}) may be  rewritten in terms of 
the Szeg\"{o} kernel (\ref{szego}) as follows:
\be
\Psi(\l_0,\l)_{kj} = S(\l^{(j)},\l_0^{(k)})E_0(\l,\l_0)
\la{psisz}
\ee
where $E_0(\l,\l_0)=(\l-\l_0)/{\sqrt{d\l} \sqrt{d\l_0}}$ is the prime-form on $\CP1$.
\end{remark}

The solution (\ref{psinew}) of the Riemann-Hilbert problem satisfies the equation (\ref{eql})
with some matrices $A_j$. Below we shall give compact expressions for the residues $A_j$.
Now we  write down a formula for $\Psi_\l\Psi^{-1}$ using a simple procedure of inversion of matrix $\Psi$. Namely, if as before we explicitly indicate dependence of matrix $\Psi$ on the
argument $\l$ and the normalization point $\l_0$ i.e. we write it as $\Psi(\l_0,\l)$,
then for an arbitrary set of three points $\l$, $\mu$ and $\nu$,
we have the well-known relation:
\be
\Psi(\mu,\l)\Psi(\l,\nu)=\Psi(\mu,\nu)\;.
\ee
In particular, for  $\mu=\nu=\l_0$, we get
\be
\Psi^{-1}(\l_0,\l)=\Psi(\l,\l_0)\;;
\la{inverse}
\ee
this relation in our case can be also verified via Fay's identity.
Therefore, we have 
\be 
\left(\Psi_\l\Psi^{-1}\right)_{lj}=\sum_{k=1}^N \psi_{\l}(\l^{(k)},\l_0^{(l)})
\psi(\l_0^{(j)}, \l^{(k)})\;,
\la{PPl}
\ee
where $\psi$ is given by (\ref{psismall}), (\ref{phinew}).
It is easy to see directly, using the formulas for $\psi(P,Q)$, that this expression has simple poles 
at all $\l_m$. Consider, for example, the contribution of a branch point $P_m$. 
In a neighbourhood of $P_m$ we have 
\be
\psi_{\l} (P,Q) \psi(Q,P) = \f{r^{(j)}_m}{\l-\l_m} (1+O(x_m)) + \f{1}{\l-\l_m}x_m^{1-\bk_m}(a_o+a_1 x_m+\dots)\;,
\la{psilpsi}
\ee
as $P\to P_m$,
where $r^{(j)}_m$ is a constant corresponding to the branch point $P_m$.
Taking into account that $\sum_{s=0}^{\bk_m-1}\g_m^{s n}=0$, where $\g_m=e^{2\pi i/\bk_m}$,
 for any $n=1,\dots,\bk_m-1$, we conclude that $\Psi_\l\Psi^{-1}$ has indeed a simple pole at $\l=\l_m$.

\section{Isomonodromic deformations and solutions of Schlesinger system}

If we now assume that vectors $\pb$, $\qb$ and constants $r_m^{(j)}$ don't depend on $\{\l_m\}$  then the monodromy matrices $M_j$ also
don't carry any $\{\l_m\}$-dependence and the   isomonodromy deformation equations take place.

\begin{theorem}
Assume that  vectors $\pb$ and $\qb$ and constants $r_m^{(j)}$   don't depend on $\{\l_m\}$. Then the
functions 
\be
A_n(\{\l_m\}) := {\rm res}|_{\l=\l_n} \left\{\Psi_{\l}\Psi^{-1}\right\}\;,
\la{solA}\ee
where  $\Psi(\l)$ is defined in (\ref{psinew}), 
satisfy the Schlesinger system (\ref{Schl}) outside of the hyperplanes
$\l_n=\l_m$ and a submanifold $(\vartheta)$ of codimension one in the $\{\l_m\}$-space defined by the condition
\be
\B\pb+\qb+\O\in (\Th)\;,
\la{Malgra}
\ee
 where $(\Th)$ denotes the  theta-divisor on Jacobian $J(\L)$.
\end{theorem}

{\it Proof.} 
We can verify validity of deformation equations (\ref{eqln}) directly
for any choice of $r_m^{(j)}$ (as long as the set $\{\l_m, {\cal M}_m\}$  
stays away from the divisor (\ref{Malgra})).
One way of proving (\ref{eqln}) is the direct computation which uses an expression for
$\Psi_{\l_m}\Psi^{-1}$ looking like (\ref{PPl}), where derivative with respect to $\l$
is substituted by derivative with respect to $\l_m$. The analysis of behaviour of 
$\Psi_{\l_m}\Psi^{-1}$ in
 a neighbourhood of $\l_m$ is then parallel to analysis of $\Psi_{\l}\Psi^{-1}$ near this point;
it shows that $\Psi_{\l_m}\Psi^{-1}$ has a simple pole at $\l_m$ with the residue equal to 
$-A_m$, according to (\ref{eqln}). Non-singularity of $\Psi_{\l_m}\Psi^{-1}$ at all other points 
$\l_n$ for $n\neq m$ can be shown analogously, which leads to
(\ref{eqln}).

Another, and simpler way to prove the deformation equations (\ref{eqln}) is to choose $\l_0=\infty$,
and use the fact that $\Psi_{\l_m}\Psi^{-1}$ is singular only at $\l_m$.
The function $\Psi$ 
is obviously invariant with respect to simultaneous shift of all $\l_m$ and $\l$ by small constant $\epsilon$: 
$\Psi(\l+\epsilon,\{\l_m+\epsilon\})=\Psi(\l,\{\l_m\})$ (this is true only if the normalization point is taken to be 
$\infty$, otherwise we have to shift $\l_0$, too).
Differentiating this relation with respect to $\epsilon$ at $\epsilon=0$, we get
$$
\Psi_\l+\Psi_{\l_1}+\dots+\Psi_{\l_M}=0\;,
$$
which implies (\ref{eqln}) with $\l_0=\infty$. Then equations
(\ref{eqln}) with arbitrary $\l_0$ are obtained by gauge
transformation of function $\Psi$.

$\Box$

Introduce the function 
\be
\hat{s}(P,Q)=\frac{\Sh(P,Q)}{\sqrt{d x_P}\sqrt{d x_Q}}
\la{shPQ}
\ee 
where the modified Szeg\"o kernel $\Sh(P,Q)$ is defined by (\ref{phinew}),
$x_P$ and $x_Q$ are local parameters at points $P$ and $Q$ respectively.
If $\l$ is used as local coordinate near $P$ and $Q$ (i.e. $P$ and $Q$  don't coincide with branch points and points at infinity), then the function $\hat{s}$ is related to the function $\psi$ (\ref{psismall})
as follows:
\be
\hat{s}(P,Q)= \f{\psi(P,Q)}{\l-\mu}\;,
\la{psihats}
\ee
where $\l=\Pi(P)$, $\mu=\Pi(Q)$.

The next proposition gives compact expressions for the solutions $\{A_m\}$ of the Schlesinger system.

\begin{theorem}\la{Amteo}
The solutions (\ref{solA}) of the Schlesinger system (\ref{Schl}) 
can be expressed as follows:
\be
(A_m)_{kj}= (\l_0-\l_m)^2 \p_{\l_m}\{\hat{s}(\l_0^{(j)}, \l_0^{(k)})\}\;,\hskip0.7cm j\neq k
\la{offdia}
\ee
\be
(A_m)_{kk}= (\l_0-\l_m)^2 \p_{\l_m} \{a_0^{(k)}\}\;,
\la{ondia}
\ee
where function $\hat{s}(P,Q)$ is defined by (\ref{shPQ}) with $\l$ used as local parameter  in a neighbourhood of
$\l_0^{(j)}$ on every sheet; $a_0^{(k)}$ is defined as a coefficient in the Laurent series:
\be
\hat{s}(\l^{(k)},\l_0^{(k)})= \f{1}{\l-\l_0} + a_0^{(k)}+ O(\l-\l_0)\;,
\hskip0.4cm{\rm as} \hskip0.4cm\l\to\l_0\;.
\la{defa0}
\ee
 \end{theorem}

{\it Proof.} From (\ref{eqln}) we find 
\ben
A_m= (\l_0-\l_m)^2 \left(\Psi_{\l_m}\Psi^{-1}\right)_\l\Big|_{\l=\l_0}\;,
\een
which implies (since $\Psi(\l_0,\l_0)=I$) that
\ben
A_m= (\l_0-\l_m)^2 \Psi_{\l\l_m}\Big|_{\l=\l_0}\;.
\een
This relation immediately leads to (\ref{ondia}), (\ref{offdia}) if we use the expression of $\Psi$ 
(\ref{psinew}) and the link (\ref{psihats}) between $\psi$ and $\hat{s}$.

$\Box$

The derivative with respect to $\l_m$ in  (\ref{offdia}), (\ref{ondia}) can  be computed using variational formulas
for all ingredients of function $\psi$ (\ref{psismall}). In general (for arbitrary multiplicities of 
branch points and non-vanishing $r_m^{(j)}$) the result turns out to be rather 
complicated. Therefore, we write the final formulas only in the following 
partial case.

\begin{corollary}
Assume that all 
branch points are simple and have different projections on $\l$-plane, and that all constants
$r_m^{(j)}$ vanish.
Then solution (\ref{offdia}),  (\ref{ondia}) of the Schlesinger system
(\ref{Schl}) can be written as follows:
\be
(A_m)_{kj}=
\f{1}{4}\left\{D_m[s(\l_0^{(j)},P_m)]s(P_m,\l_0^{(k)})-s(\l_0^{(j)},P_m)
D_m[s(P_m,\l_0^{(k)})] \right\}
\ee
for any $j,k=1,\dots,N$, where $s(P,Q)$ is given by (\ref{szego}), (\ref{sPQ}).
\end{corollary}

\begin{remark}\rm
Our solutions $\{A_m (\{\l_n\})\}$ (\ref{offdia}), (\ref{ondia})
 of the Schlesinger system is singular on the
Malgrange divisor $(\vartheta)$ (it has a pole whose order is equal to
 the order of the zero of the tau-function at the point of the divisor). 
Since this divisor carries a non-trivial dependence on monodromy matrices,
which parametrise the initial conditions for  the Schlesinger system, this singularity depends on 
initial conditions i.e. it is ``movable''.
\end{remark}

The next section is devoted to computation of tau-function
(\ref{taudef}) corresponding to solutions of Schlesinger system given
by the theorem \ref{Amteo}.

\section{Isomonodromic tau-function}
\la{sectau}

\subsection{Tau-function and Bergmann projective connection}

According to the definition of the tau-function (\ref{taudef}), let us start with calculation of expression
$\tr\left(\Psi_\l\Psi^{-1}\right)^2$.
Notice that this object is independent of the choice of
normalization point $\l_0$ [substitution of $\l_0$ by another point $\tilde{\l}_0$ corresponds to the
$\l$-independent ``gauge'' transformation $\Psi(\l)\to \tilde{\Psi}(\l)= \Psi^{-1}(\tilde{\l}_0)\Psi(\l)$].

 Consider the limit $\l_0\to\l$ in the formula (\ref{psinew}) for
$\Psi_{jk}$, where $\Sh(P,Q)$ is given by expression (\ref{phinew}). In this limit  matrix elements of the function $\Psi$ behave as follows:
\be
\Psi_{kj}(\l,\l_0)= \f{\l_0-\l}{d\l}\Sh(\l^{(j)},\l^{(k)})  + O\{(\l_0-\l)^2\}\;,\hskip0.6cm k\neq j
\ee
\be
\Psi_{jj}(\l,\l_0)= 1 + \f{\l_0-\l}{d\l}\left\{W_1(\l^{(j)}) - W_2(\l^{(j)})\right\}\;,
\ee  
where $W_1(P)$ is a linear combination of the basic   holomorphic 1-forms on $\L$:
\be
W_1(P) = \f{1}{\Thpq(\O)} \sum_{\a=1}^g \p_{z_\a}\{\Thpq(\O)\} w_\a (P)\;,
\la{W1}\ee
and $W_2(P)$ is the following meromorphic 1-form with simple poles at the points $\l_m^{(j)}$ and
the residues $r_m^{(j)}$:
\be
W_2(P)=  \sum_{m=1}^M\sum_{j=1}^N r_m^{(j)}d_P
\log E(P,\l_m^{(j)})\;.
\ee

Taking into account independence of the expression $\tr\left(\Psi_\l\Psi^{-1}\right)^2$  on   position of the normalization point 
$\l_0$, we have
\ben
\tr\left(\Psi_\l\Psi^{-1}\right)^2(d\l)^2=
2\sum_{j<k}\Sh(\l^{(j)},\l^{(k)})\Sh(\l^{(k)},\l^{(j)}) + \sum_{j=1}^N \left(W_1(\l^{(j)})-
W_2(\l^{(j)})\right)^2\;.
\een
To transform this expression  we first notice that, according to (\ref{SzBerg}),
\ben
\Sh(P,Q)\Sh(Q,P)= -\w(P,Q)-\sum_{\a,\b=1}^g\p^2_{z_\a z_\b}\{\log\Thpq(\O)\}w_\a (P) w_\b(Q)\;.
\een
Furthermore, since $W_1(P)$ is a holomorphic 1-form on $\L$, the expression 
$\sum_{j=1}^N W_1(\l^{(j)})$ vanishes identically according to Lemma \ref{holsum}; hence
\ben
\sum_{j=1}^N \{W_1(\l^{(j)})\}^2 
= -2\sum_{\stackrel{j,k=1}{j<k}}^N \sum_{\a,\b=1}^g\p_{z_\a}\{\log\Thpq(\O)\}
\p_{z_\b}\{\log\Thpq(\O)\} w_\a(\l^{(j)})w_\b(\l^{(k)})\;.
\een

Similarly, we can conclude that $\sum_{j=1}^N \{W_2(\l^{(j)})\}^2$ is a meromorphic 2-form on $\CP1$ which has
poles only at the points $\l_m$; calculation of corresponding residues gives
\be
\sum_{j=1}^N \{W_2(\l^{(j)})\}^2=\sum_{m,n=1}^M \f{r_{mn}(d\l)^2}{(\l-\l_n)(\l-\l_m)}\;,
\ee
where
\be
r_{mn}=\sum_{j=1}^N r_m^{(j)} r_n^{(j)}\;.
\la{rmn}
\ee

Therefore, as the first step of our calculation,  we get the following expression:
\be
\f{1}{2}\tr\left(\Psi_\l\Psi^{-1}\right)^2(d\l)^2=
-\sum_{j<k} \w(\l^{(j)},\l^{(k)})
\la{trace}
\ee
\ben
-\f{1}{\Thpq(\O)}\sum_{j< k}\sum_{\a,\b} \p^2_{z_\a z_\b}\{\Thpq(\O)\} w_\a(\l^{(j)})
w_\b(\l^{(k)})
+
\f{1}{2}\sum_{m,n}\f{r_{mn}(d\l)^2}{(\l-\l_n)(\l-\l_m)}
\een
\ben
-\f{1}{\Thpq(\O)}\sum_{\a}\p_{z_\a}\{\Thpq(\O)\}\sum_{m}\sum_{j} r_m^{(j)}w_\a(\l^{(j)})
d_P \log E(P,\l_m^{(j)})\;.
\een

Let us now analyze the Hamiltonians 
$$
H_m\equiv \frac{1}{2}{\rm res}|_{\l=\l_m} \left\{\tr\left(\Psi_\l\Psi^{-1}\right)^2\right\}\;
$$
(to avoid confusion we notice that in this section we use only the notion of residue of {\it function}
of variable $\lambda$ at finite points of the complex plane). 
Using the heat equation for theta-function (\ref{heat}), and Rauch's
 formula (\ref{varB1}),
we can represent $H_m$  in the following form:
\be
H_m = -\res|_{\l=
\l_m}\f{1}{(d\l)^2}\left\{\sum_{j<k} \w(\l^{(j)},\l^{(k)}) \right\}+ \f{1}{2}\sum_{n\neq m}
\f{r_{mn}}{\l_m-\l_n}
\la{Hm}\ee
\ben
+\f{1}{\Thpq(\O)}\sum_{\a,\b}\f{\p\Thpq(\O)}{\p\B_{\a\b}} \p_{\l_m}
\{\B_{\a\b}\}+\f{1}{\Thpq(\O)}\sum_{\a}\p_{z_\a} \{\Thpq(\O)\}\p_{\l_m}\{\O_\a\}\;,
\een
or, equivalently,
\be
H_m= -\res|_{\l=\l_m}\left\{\f{1}{(d\l)^2}\sum_{j<k} \w(\l^{(j)},\l^{(k)})\right\}
+ 
\p_{\l_m}\log\left\{\prod_{l<n}(\l_l-\l_n)^{r_{ln}}\Thpq(\O)\right\}\,.
\la{Hm10}
\ee
Therefore, we come to  the following 
\begin{theorem}
The tau-function
corresponding to solution (\ref{solA}) of Schlesinger system, is given by
\be
\tau(\{\l_n\}) = F(\{\l_n\}) \prod_{m,n=1}^M (\l_m-\l_n)^{r_{mn}}\Thpq\left(\O|\B\right)\;,
\la{tau00}\ee
where function $F(\{\l_n\})$ does not depend on constants $\pb,\qb$ and
 $r_n^{(j)}$, and satisfies the following system of compatible equations
\be
\p_{\l_m}\{\log F\} = \sum_{P_l\in \Pcal_m}\f{1}{12\bk_l}\f{R_l^{(\bk_l-2)}(P_l)}{(\bk_l-2)!}\;,
\la{wRH}\ee
where $R_l(P)$ is Bergmann projective connection corresponding to our  choice
of system of local parameters near branch points $P_l\in\Pcal_m$: 
$\lp_l=(\l-\l_m)^{1/\bk_l}$. 
\end{theorem}

{\it Proof.} According to the expressions (\ref{Hm10}), we need to
prove the following lemma.

\begin{lemma}
The following identity holds:
\be
\sum_{P_l\in \Pcal_m}\f{1}{12\bk_l}\f{R_l^{(\bk_l-2)}(P_l)}{(\bk_l-2)!}= -\res|_{\l=\l_m}\left\{
\sum_{j\neq k}\frac{\w(\l^{(j)},\l^{(k)})}{(d\l)^2}\right\}\;.
\la{F1}\ee
\end{lemma}
{\it Proof.} The right hand side of the formula (\ref{F1}) can be rewritten in terms of non-singular part  $H$
of the
Bergmann kernel    (\ref{defH}) as follows: 
\be
-\res|_{\l=\l_m}\sum_{P_l\in\Pcal_m}\left\{\sum_{j,k=1}^{\bk_l}
H(\g_l^j\lp_l,\g_l^k\lp_l)\g_l^{j+k} 
\left(\f{d\lp_l}{d\l}\right)^2\right\}\;,
\la{minres}
\ee
where $\g_l=exp\f{2\pi i}{\bk_l}$.
In terms of coefficients of the  Taylor series of $H(\lp_l, y_l)$
around $P_l$ we have:
\ben
H(\lp_l,y_l)=\sum_{s=0}^{\infty}\sum_{p=0}^s \f{H^{(p,s-p)}(0,0)}{p!
(s-p)!}\lp_l^p y_l^{s-p}\;,
\een
and  expression (\ref{minres}) looks as follows:
\ben
-\sum_{P_l\in\Pcal_m}\f{1}{\bk_l^2}\sum_{p=0}^{\bk_l-2} 
\f{H^{(p,\bk_l-2-p)}(0,0)}{p! (\bk_l-2-p)!}\sum_{j,k=1\,,\; j< k}^{\bk_l}
\g^{(p+1)k + (\bk_l-p-1)j}\;.
\een
Summing up the geometrical progression, we get:
\ben
\sum_{P_l\in\Pcal_m}\f{1}{2\bk_l}\sum_{p=0}^{\bk_l-2} 
\f{H^{(p,\bk_l-2-p)}(0,0)}{p! (\bk_l-2-p)!}=
\sum_{P_l\in\Pcal_m}\f{1}{12\bk_l}\f{R_l^{(\bk_l-2)}(0)}{(\bk_l-2)!}\;.
\een
$\Box$

$\Box$

One can check that only the non-singular part of the Bergmann kernel
contributes to the residue in (\ref{F1});  therefore,
we can further express $\p_{\l_m}\log F$ in terms of the Bergmann
projective connection  corresponding to
the natural choice of local coordinates on $\L$ on the 
branched covering $\L$.

\begin{remark}\rm
Suppose that all projections of branch points   $P_m$ on $\CP1$ are
different. Then
equations (\ref{wRH}) for function  $F(\{\l_n\})$ can be written as follows: 
\be
\p_{\l_m}\{\log F\} = 
-\f{1}{12\pi}\int_{\L}  \mu_m R_m (d\lp)^2\;,
\la{wRH1}
\ee
where $\mu_m$
is the Beltrami differential (\ref{schif}) corresponding to variation
of the branch point $P_m$.
\end{remark}

\begin{theorem}
The following equations for Bergmann projective connection on the branch
covering $\L$ are fulfilled:
\ben
\f{\p}{\p\l_n}\left\{\sum_{P_l\in {\cal P}_m}\frac{1}{(\bk_l-2)!\,\bk_l}
\left(\frac{d}{d\lp_l}\right)^{\bk_l-2}R_l(\lp_l)\big|_{\lp_l=0}\right\}
\een
\be
=\f{\p}{\p\l_m}\left\{
\sum_{P_l\in {\cal P}_n}\frac{1}{(\bk_l-2)!\,\bk_l}
\left(\frac{d}{d\lp_l}\right)^{\bk_l-2}R_l(\lp_l)\big|_{\lp_l=0}\right\}\;.
\la{compint1}\ee
\end{theorem}

{\it Proof.} Equations (\ref{compint1}) provide 
integrability of equations (\ref{wRH}) for the function $F$ which follows
from integrability of the equations (\ref{taudef}) for the
isomonodromic tau-function.

$\Box$

\begin{corollary}
Let all branch points of $\L$ be  simple and have
different projections on $\l$-plane. Then  values of Bergmann
projective connection computed with respect to the natural system of
local parameters on $\L$ (i.e. $\lp_m=\sqrt{\l-\l_m}$ at the branch
point $P_m$) satisfy the following equations: 
\be
\f{\p R_m(P_m)}{\p\l_n}=
\f{\p R_n(P_n)}{\p\l_m}\;.
\la{compat}\ee 
\end{corollary}

These equations  are analogous to  equations for accessory parameters which appear
in the uniformization problem of punctured sphere \cite{TakZog93}.

Since the Bergmann projective connection $R_m(P_m)$ is finite and
holomorphic function of $\{\l_m\}$ as long as the Riemann surface $\L$
remains non-degenerate, we conclude that 
the function $F$ does not vanish and remains finite outside of the hyperplanes $\l_m=\l_n$. This allows to claim that the divisor of  zeros of the tau-function
(\ref{tau00}) coincides with the divisor of  zeros of the theta-function $\Thpq\left(\O|\B\right)$:  
\begin{theorem}
The set of singularities   $\{\l_m\}$ lies in  the Malgrange divisor
$(\vartheta)\subset \C^M$ iff the vector $\B\pb+\qb+\O$ belongs to the theta-divisor $(\Th)$ 
in  the Jacobi manifold $J(\L)$ of the Riemann surface $\L$.
\end{theorem}  
We  remind that in the  expression $\B\pb+\qb+\O$ the $\{\l_m\}$-dependence is hidden inside the  matrix of $b$-periods and the vector $\O$.

\begin{remark}\rm
It turns out \cite{KokKor03} that function $F$ itself coincides with the isomonodromic tau-function of 
another  RH problem introduced by Dubrovin \cite{Dub94} in the context of Frobenius manifolds 
associated with Hurwitz spaces. It would be interesting to obtain the explicit link between this
RH problem and the one studied in this paper on the level of monodromy representations
and function $\Psi$. 
\end{remark}

\subsection{Riemann-Hilbert
problems with off-diagonal $2\times 2$  monodromy matrices }

Here we consider the simplest
case of $N=2$, when any matrix of quasi-permutation is either diagonal
or off-diagonal. 
We shall consider  monodromy groups such that 
all monodromies $M_m$ are  off-diagonal; the insertion of additional diagonal monodromies according to the general scheme is straightforward.
In this case the branched covering $\L$ corresponds to hyperelliptic algebraic curve with branch points $\l_1,\dots,\l_M$ and function $F$ 
may be calculated explicitly \cite{KitKor98}.
We have  $M=2g+2$, where  $g$ is the genus of the hyperelliptic curve  $\L$: 
\be
w^2=\prod_{m=1}^{2g+2}(\l-\l_m)\;.
\la{he}\ee

Let us put all $r_m^{(j)}=0$;  in this case the  formula (\ref{psisz}) gives the  solution $\Psi(\l)\in SL(2)$ of the RH problem with
arbitrary off-diagonal $SL(2)$-valued monodromies:

\ben
M_m=\left(\ba{cc}                0        &  d_m      \\
                        -d_m^{-1}    &     0        \ea\right) \;,
\een
where constants $d_m$ may be expressed in terms of the  elements of vectors $\pb,\qb$. Let us count the 
number of essential parameters in
the monodromy matrices and in the construction of function $\Psi$. The matrices $M_m$ contain altogether 
$2g+2$ constants; however,
there is one relation (product of all monodromies gives $I$). One more parameter is non-essential due to 
possibility of simultaneous
conjugation of all monodromies with an arbitrary  diagonal constant matrix.
Therefore, the set of monodromy matrices contains $2g$ non-trivial constants in accordance with the 
number of non-trivial constants contained 
in the vectors $\pb$ and $\qb$.

To integrate the remaining equations
\be
\p_{\l_m}\log F = \f{1}{24}R(\l_m)
\la{FHE}
\ee
 on  hyperelliptic curve (\ref{he}) we use the following formula (\cite{Fay73}, p.20) 
for the Bergmann  projective connection at arbitrary point of  $P$ of the
hyperelliptic curve $\L$  (where $\lp$ is the local parameter in the neighborhood of the point $P$, $\l=\Pi(P)$ is the projection of $P$ on $\l$-plane):
\be
R(P)=\{\l(\lp),\lp\}(P)+\f{3}{8}\left(\f{d}{d \lp}\log\f{\prod_{\l_m\in T}(\l-\l_m)}{\prod_{\l_m\not\in T}(\l-\l_m)}\right)^2 (P)
\la{prhyp}\ee
\ben
-\f{6}{\Th\left[^{\pb^T}_{\qb^T}\right](0)}\sum_{\a,\b=1}^g \p^2_{z_\a z_\b}\left\{\Th\left[^{\pb^T}_{\qb^T}\right](0)\right\}
\f{w_\a}{d \lp}(P)\f{w_\b}{d \lp}(P)\;.
\een
Here $\{\l,\lp\}$ is the Schwarzian derivative of $\l$ with respect to $\lp$; 
$T$ is an arbitrary divisor consisting of  $g+1$ branch points, which satisfies certain non-degeneracy condition. Characteristic
$\left[^{\pb^T}_{\qb^T}\right]$ is the even half-integer  characteristic corresponding to the divisor
$T$  according to  the following equation:
\be
\B\pb^T+\qb^T=\sum_{\l_m\in T} U(\l_m)-\Delta\;,
\la{pqT}\ee
where $\Delta$ is a  vector of Riemann constants; the initial point of the Abel map is chosen to be, say, $\l_1$. In this case the
r.h.s. of (\ref{pqT}) is a linear combination, with integer or half-integer coefficients, of the  vectors $\eb_\a$ and $\B\eb_\a$. These  coefficients
are composed in vectors $\pb^T$ and $\qb^T$. The non-degeneracy  requirement imposed on the divisor $T$ gives rise to the condition that
the vector  $\B\pb^T+\qb^T$
does not belong to the  theta-divisor on $J(\L)$, i.e.
 $\Th\left[^{\pb^T}_{\qb^T}\right](0)\neq 0$.

The Bergmann projective connection $R$, as well as the function $F$,  are independent of the  choice of the divisor $T$.
If in (\ref{prhyp}) we choose $P=\l_m$, the local parameter is $\lp=\sqrt{\l-\l_m}$. Then all terms in $R_m(\l_m)$
which don't contain theta-function can be integrated explicitly; the terms
 containing theta-function
can be represented as logarithmic derivative with respect to $\l_m$ by making  use of  the heat equation for 
theta-function (\ref{heat}) and Rauch formula (\ref{varB1}). These terms are equal to 
$-6 \p_{\l_m}\log\Th\left[^{\pb^T}_{\qb^T}\right](0)$.
In turn, this expression may be rewritten using the Thomae formula 
\ben
\Th\left[^{\pb^T}_{\qb^T}\right](0)=\pm(\det{\cal A})^2 \prod_{\l_m,\l_n\in T} (\l_{m}-\l_{n})
 \prod_{\l_m,\l_n\not\in T}(\l_{m}-\l_{n}),
\een
where ${\cal A}_{\a_\b}=\oint_{a_\a}\f{\l^{\b-1}}{w}$ is the $g\times g$ matrix of $a$-periods of non-normalized holomorphic differentials on $\L$.
Collecting all  factors arising from the Thomae formula and from the expression (\ref{prhyp}), we get the following answer for the
function $F$:
\be
F=[\det{\cal A}]^{-\f 12}
\prod\limits_{m<n}(\l_m-\l_n)^{-\frac 18}\;,
\la{Fhe}
\ee
which is equal to   $\{\det\dbar_0\}^{-1/2}$, where $\det\dbar_0$ can be interpreted
as (defined and computed heuristically \cite{Zamo86,Kniz87})  determinant of Cauchy-Riemann operator acting in trivial bundle over $\L$. For the tau-function itself we get the following expression
\ben
\tau(\{\l_m\})=[\det{\cal A}]^{-\f 12}
\prod\limits_{m<n}(\l_m-\l_n)^{-\frac 18}\Theta\left[^\pb_\qb\right](0|\B)\;,
\een
which can be interpreted as   determinant of the Cauchy-Riemann operator acting on 
the sections of line bundle $\Delta\otimes\chi_{\pb,\qb}$ over $\L$.

\subsection{Tau-function in genus 0 and 1}

The function $F$  (\ref{F1}), and, therefore, the Jimbo-Miwa
tau-function  can be also calculated for monodromy groups
corresponding to arbitrary
coverings of genus $0$ and $1$. 

\begin{theorem}
Let the branched covering $\L$ corresponding to a Riemann-Hilbert
problem with quasi-permutation monodromies have genus 0. Then the 
Jimbo-Miwa tau-function has the following form:
\be
\tau = \left\{\frac{\prod_{m=1}^M(\frac{dU}{dx_m}(P_m))^{\frac{\bk_m-1}{2}}}
{\prod_{k=2}^N(\frac{dU}{d\zeta_k}(\infty^{(k)})}\right\}^{1/12} \prod_{m,n=1}^M (\l_m-\l_n)^{r_{mn}}\;,
\la{genus0}
\ee
where $U(P): \L\to \CP1$ is the uniformization map of the branched
covering $\L$ fixed by the condition $U(\infty^{(1)})=\infty$;
$\lp_m=(\l-\l_m)^{1/\bk_m}$ are local parameters near the branch
points; $\zeta_k\equiv 1/\l$ are the local parameters around
infinities of $\L$.
  Constants $r_{mn}$ are given by (\ref{rmn}).
\end{theorem}

Analogous statement is valid in genus 1 case:

\begin{theorem}
Let the branched covering $\L$ corresponding to a Riemann-Hilbert
problem with quasi-permutation monodromies have genus 1. Then the 
Jimbo-Miwa tau-function has the following form:
\be
\tau = \left\{\frac{\prod_{m=1}^M(\frac{dU}{dx_m}(P_m))^{\frac{\bk_m-1}{2}}}
{\prod_{k=1}^N(\frac{dU}{d\zeta_k}(\infty^{(k)})}\right\}^{1/12}\prod_{m,n=1}^M (\l_m-\l_n)^{r_{mn}} 
\f{\Thpq(\O|\mu)}{[\Theta_1'(0|\mu)]^{1/3}}\;,
\la{genus1}
\ee
where $U(P)=\int^P w(P)$ is the uniformization map of the branched
covering $\L$ to its fundamental parallelogramm with periods $1$ and
$\mu$ ($w(P)$ is the normalized
holomorphic 1-form on $\L$); $\theta_1$ is the odd Jacobi
theta-function on $\L$;
$\lp_m=(\l-\l_m)^{1/\bk_m}$ are local parameters near the branch
points; $\zeta_k\equiv 1/\l$ are the local parameters around
infinities of $\L$.
  Constants $r_{mn}$ are given by (\ref{rmn}); argument $\O$ of the
theta-function is defined by  (\ref{Om}).
\end{theorem}

The proofs of both theorems are contained in \cite{KokKor02}; they
are based on the properties of Dirichlet  action corresponding to
the  metric  $d\l d\bar{\l}$ in genus zero, and the metric
$w\bar{w}$ in genus 1.


{\bf Acknowledgements.}
This work was strongly influenced by Andrej Nikolaevich Tyurin.
I would like to thank Alexandr Bobenko, John Harnad, Jacques Hurtubise, Alexey Kokotov,
Vladimir Matveev and Alexandr Orlov for important discussions. Several suggestions 
of anonymous referee
were used in the final version of this work. 
This work was supported by  NSERC grant, FCAR grant and  FRDP grant of
Concordia University.

\end{document}